\documentclass[aps,prl,twocolumn,showpacs,superscriptaddress,groupedaddress]{revtex4}  
\usepackage{graphicx}  
\usepackage{dcolumn}   
\usepackage{bm}        
\usepackage{amssymb}   
\usepackage{mciteplus}

\begin{document}



	\title{Thermodynamic of a rotating and Non-linear magnetic-charged black hole in the quintessence field}

\thanks{Corresponding author: nragilbrand@gmail.com}%

\author{R. T. Ndongmo}
\affiliation{%
	Department of Physics, Faculty of Science, University of Yaounde I, P.O. Box. 812, Yaounde, Cameroon, \\
}

\author{S. Mahamat}
\affiliation{
	Department of Physics, Higher Teacher’s Training College,  University of Maroua, P.O. Box 55, Maroua, Cameroon,\\
}

\author{T. B. Bouetou}
\affiliation{%
	Department of Physics, Faculty of Science, University of Yaounde I, P.O. Box. 812, Yaounde, Cameroon, \\
}
\affiliation{
	National Advanced School of Engineering, University of Yaounde I, P.O. Box. 8390, Yaounde,
	Cameroon \\
}%

\author{T. C. Kofane}
\affiliation{%
	Department of Physics, Faculty of Science, University of Yaounde I, P.O. Box. 812, Yaounde, Cameroon, \\
}
\affiliation{%
	The African Center of Excellence in Information and Communication Technologies, Yaounde,	Cameroon
}%


\date{\today}

\begin{abstract}
	
\end{abstract}

\begin{abstract}
	We purpose an approach for the thermodynamic analysis of rotating and non-linear magnetic-charged black hole with quintessence. Accordingly, we compute various thermodynamics quantities of the black hole, such as mass, temperature, potential provided from the magnetic charge, and the heat capacity. Moreover, we study phase transitions of this black hole, analyzing the plot of its heat capacity. Then, we have shown that the black hole mass would have a phase of decrease, while the temperature increases from negative absolute temperatures. From the behavior of the heat capacity, we point out that the black hole undergoes to a second-order phase transition, which is shifted towards the higher values of entropy as we increase the rotating parameter $a$ or the magnetic parameter $Q$.

\end{abstract}
\maketitle

\section{I. INTRODUCTION}

The discovery in the early 1970s that black holes radiate as black bodies has radically affect our understanding of general relativity, and offer us some early hints about the nature of quantum gravity. As a result, black holes may behave as thermodynamic objects, with characteristic temperature and entropy. So, black holes thermodynamics started since the seminal works of Hawking\cite{Hawking1974} and Bekenstein\cite{Bekeinstein1973}. Moreover, through an experimental way, Steinhauer et al.\cite{steinhauer2014observation,steinhauer2016observation,de2019observation} have observed spontaneous Hawking radiation, stimulated by quantum vacuum fluctuations, emanating from an analogue black hole in an atomic Bose-Einstein condensate. This kind of black hole is called acoustic black hole\cite{visser1998acoustic,balbinot2008nonlocal}, because instead of trapping the light, it traps the sound.

Accelerated expansion of the universe is the most recent fascinating result of observational cosmology\cite{riess1998observational,riess1999bvri,perlmutter1999measurements}. To explain the accelerated expansion of the universe, it is proposed that the universe is regarded as being dominated by an exotic scalar field with a large negative pressure called "dark energy", which constitutes about $70\%$ of the total energy of the universe\cite{Kuriakose2013}. Thus it is the major component of the universe. There are several candidates for dark energy. Besides the cosmological constant\cite{krauss1995cosmological,weinberg1989cosmological,peebles2003cosmological,padmanabhan2003cosmological}, "Quintessence" is one among them. It is characterized by a parameter $\epsilon$, the ratio of the pressure to energy density of the dark energy, and the value of $\epsilon$ falls in the range $-1\leq \epsilon\leq -\frac{1}{3}$\cite{shahjalal2019thermodynamics,javed2019fermions,Kuriakose2013}. Kiselev \cite{Kiselev2003} first derived the solutions of the black hole surrounded by the quintessence. Since then, many black holes in the quintessence field have been developed\cite{chen2008hawking,yi2011thermodynamic,thomas2012thermodynamics,fernando2013nariai,ghaderi2016thermodynamics,li2014effects}.

Several works have been done on the black hole phase transition\cite{Banerjee2011b,appels2016thermodynamics,hennigar2017superfluid}. Already in 1983, the classic paper by Hawking and Page\cite{Hawking1983} on black hole phase transitions appeared.  The black hole phase transition can be studied theoretically in the light of the behavior of its heat capacity\cite{Kuriakose2013}. 	Husain and Mann\cite{Husain2009} suggested that the specific heat of a black hole becomes positive after a phase transition near the Planck scale. A lot of scientist have studied phase transitions caused by quintessence type of dark energy; Kamiko et al.\cite{Kamiko2018} investigated second-order thermodynamic phase transition of Regular Hayward Black hole surrounded by	quintessence, and Mahamat et al\cite{Mahamat2018}. have investigated the second-order thermodynamic phase transition of the Regular Bardeen black hole.

Recently, a new proof of existence of black holes has been shown by the Event Horizon Telescope collaboration\cite{collaborat2019first,akiyama2019first,akiyama2019first3,collaborat2019first4,akiyama2019first5}. They have presented the first image of a central supermassive black hole, named M87. On the other hand, observations have shown that these supermassive black holes are rotating\cite{kawashima2019black,murchikova2019cool}. Then, it could be necessary to carry out a theoretical study of these stars and their interactions within dark energy. Thereby, Toshmatov et \textit{al.}\cite{Toshmatov2017rotating} have founded a rotating black hole solution with quintessence field and have studied its thermodynamic behavior.

On the other hand, black holes, predicted by the Einstein theory of General Relativity, have a propriety known as Singularity, at which densities and curvature become infinite\cite{hawking2010nature}. Among the different alternatives for that, a kind of black hole with regular non-singular geometry with an event horizon satisfying weak energy condition was constructed by Bardeen\cite{Mahamat2018,ayon2000bardeen} and has been obtained introducing an energy-momentum tensor interpreted as the gravitational field of some sort of nonlinear magnetic monopole charge $Q$. Thereby, the metric of non-linear magnetic-charged black hole surrounded by quintessence has been obtained by Nam\cite{nam2018non}.

However, Benavides et \textit{al.}\cite{benavides2018rotating} have putted out the metric of rotating and non-linear magnetic-charged black hole in the quintessence field, which is the combination of both previous metrics. Their results give us many informations about the geometric of this black hole. The aim of the paper is to study the thermodynamic behavior of rotating and non-linear magnetic-charged black hole in the quintessence field.

Here, we shall compute the entropy of rotating and non-linear magnetic-charged black hole with quintessence. Afterwards, we shall compute and plot the variation of some thermodynamic quantities, such as the temperature, mass and potential provided from the magnetic charge, then we analyze effects of the magnetic charge and the rotating parameter on its thermodynamic behavior. At the end, we calculate the heat capacity  and analyze its plot, then we see how phase transitions occur in this kind of black hole.

\section{II. ENTROPY OF ROTATING AND NON-LINEAR MAGNETIC-CHARGED BLACK HOLE IN THE QUINTESSENCE FIELD}
Benavides et \textit{al.}\cite{benavides2018rotating} have derived the metrics of rotating and non-linear magnetic-charged black hole with quintessence, which is expressed as

\begin{equation}
\left.
\begin{array}{r c l}
ds^2&=&g_{\mu\nu}dx^\mu dx^\nu\\
&=&-\left[1-\frac{2\rho r}{\Sigma}\right]dt^2+\frac{\Sigma}{\Delta}dr^2-\frac{4a\rho r \sin^2\theta}{\Sigma}dtd\phi\\\\
&+& \Sigma d\theta+\sin^2\theta\left[r^2+a^2+\frac{2a^2\rho r \sin^2\theta}{\Sigma}\right]d\phi^2,
\end{array}
\right.
\end{equation}
where
\begin{equation}
\left.
\begin{array}{r c l}
\Delta&=&r^2-2\rho r+a^2,\\
\Sigma&=&r^2+a^2\cos^2\theta,\\
2\rho&=&\frac{2Mr^3}{Q^3+r^3}+\frac{c}{r^{3\epsilon}}.
\end{array}
\right.
\end{equation}

Here, the geometry of the black hole is expressed using spherical coordinates $(r,\ \theta,\ \phi)$, $Q$ is the magnetic charge, $a$ is a parameter of the rotating black hole, $\epsilon$ and $c$ are parameters of quintessence.


For a rotating black hole, the expression of entropy is found throughout the horizon area\cite{Toshmatov2017rotating}. Therefore, we have

\begin{equation}
S=\frac{A}{4}=\frac{1}{4}\int_{0}^{2\pi}d\phi\int_{0}^{\pi}d\theta\sqrt{g_{\theta\theta}g_{\phi\phi}},
\end{equation}
which gives us
\begin{eqnarray*}
	S&=&\frac{\pi}{2}\int_{0}^{\pi}d\theta\sqrt{\Sigma\sin^2\theta\left[r_h^2+a^2+\frac{2a^2\rho r_h \sin\theta}{\Sigma}\right]}\\
	&=&\pi(r_h^2+a^2).
\end{eqnarray*}

This leads us to write the expression of the horizon radius as
\begin{equation}
r_h=\sqrt{\frac{S}{\pi}-a^2}.
\end{equation}

The variation of entropy $S$ in term of $r_h$ is plotted in FIG. 1.

\begin{figure}[!h]
	\begin{center}
		\includegraphics[scale=.40]{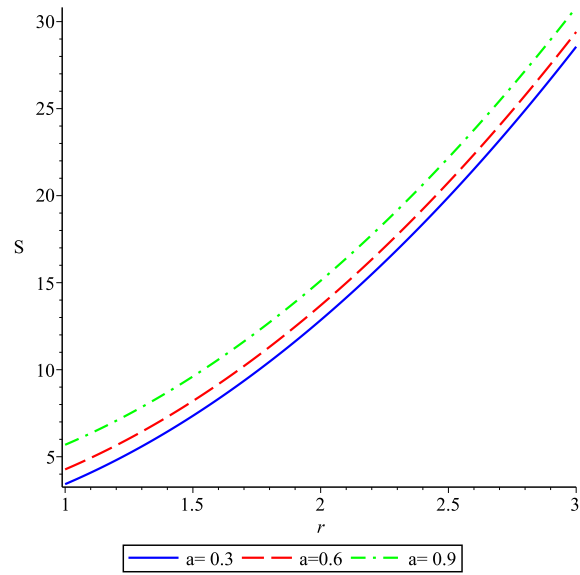}
	\end{center}
	\caption{Variation of the entropy $S$ in the presence of
		quintessence dark energy. Here $a=0.3$ corresponds to blue continuous line, $a=0.6$ to red dash and $a=0.9$ to green dash-point.}
\end{figure}
\newpage
Here, we see that whatever the value of $a$, the entropy is always growing up. This could permit us to confirm the second law of black hole thermodynamic, which tels us that entropy $S$ of black hole is always increasing.  Furthermore, we notice that for each value of the parameter $a$, entropy is higher when the horizon radius $r_h$ increases.

\section{III. PHASE TRANSITION OF THE BLACK HOLE}

\ \ \ \ \ \ \ \ \textbf{1. Thermodynamics quantities}

In the case of rotating and non-linear magnetic-carded black hole, we have the presence of event and quintessence horizons\cite{nam2018non}, In our work, we will focus on event horizon and make a thermodynamic analysis. In this case, the first law of the thermodynamic is established as\cite{smarr1973mass}
\begin{equation}
dM=TdS+\Phi dQ+Cdc+\Omega dJ.
\end{equation}
Here, $\Phi$ is the Potential provided from the magnetic charge, $C$ the thermodynamic quantity corresponding to the quintessence, $\Omega$ is the angular velocity and $J$ the angular momentum.

To compute the mass $M$, we will use the event horizon propriety\cite{benavides2018rotating}, expressed as

\begin{equation}
g^{rr}=0 \Rightarrow \ \frac{\Delta}{\Sigma}=0.
\end{equation}
This leads us to have
\begin{equation}
M=\frac{1}{2}(Q^3+r_h^3)\left(\frac{1}{r_h^2}+\frac{a^2}{r_h^4}-\frac{c}{r_h^{3\epsilon+3}}\right).
\end{equation}
When $a=0$ and $\epsilon=-\frac{2}{3}$, we obtain the same result as Nam\cite{nam2018non} expressed
\begin{equation}
M=\frac{Q^3+r^3}{2r_h^2}\left(1-cr_h\right).
\end{equation}

In term of $S$, the mass $M$ gets the form
\begin{widetext}
	\begin{equation}
	 M=\frac{1}{2}\left(Q^3+\left(\frac{S}{\pi}-a^2\right)^\frac{3}{2}\right)\left(\frac{1}{\left(\frac{S}{\pi}-a^2\right)}+\frac{a^2}{\left(\frac{S}{\pi}-a^2\right)^2}-\frac{c}{\left(\frac{S}{\pi}-a^2\right)^\frac{3\epsilon+3}{2}}\right).
	\end{equation}
\end{widetext}
Now, let us compute the temperature $T$,  the potential $\Phi$ and the quantity $C$. their formulas are respectively
\begin{eqnarray}
T&=&\left(\frac{\partial M}{\partial S}\right)_{Q,c,J}\\
\Phi&=&\left(\frac{\partial M}{\partial Q}\right)_{S,c,J}\\
C&=&\left(\frac{\partial M}{\partial c}\right)_{S,Q,J}
\end{eqnarray}

Developing Eq.(10) we get

\begin{widetext}
	
	\begin{eqnarray}
	\left.
	\begin{array}{r c l}
	 T&=&\frac{3}{4}\frac{\sqrt{\frac{S}{\pi}-a^2}\left(\frac{1}{\frac{S}{\pi}-a^2}+\frac{a^2}{\left(\frac{S}{\pi}-a^2\right)^2}-\frac{c}{\left(\frac{S}{\pi}-a^2\right)^{\frac{3\epsilon+3}{2}}}\right)}{\pi}\\\\
	 &+&\frac{1}{2}\left(Q^3+\left(\frac{S}{\pi}-a^2\right)^\frac{3}{2}\right)\left(-\frac{1}{\pi\left(\frac{S}{\pi}-a^2\right)^2}-\frac{2a^2}{\pi\left(\frac{S}{\pi}-a^2\right)^3}+\frac{c\left(\frac{3\epsilon+3}{2}\right)}{\pi\left(\frac{S}{\pi}-a^2\right)^\frac{3\epsilon+5}{2}}\right).
	\end{array}
	\right.
	\end{eqnarray}
	
\end{widetext}

The behavior of the temperature is plotted on Fig. 2. Through that figure, we can see that the temperature $T$ starts increasing, from absolute negative values to positives values and decreases towards zero. So, we could notice that rotating parameter $a$ allows temperature to have negative values, then get positive values. In addition, the plot of the temperature shows us that as we increase the value of $a$, the maximum of temperature shifted to lower values of the temperature and higher values of the entropy.

\begin{figure}[!h]
	\centering
	\includegraphics[scale=0.35]{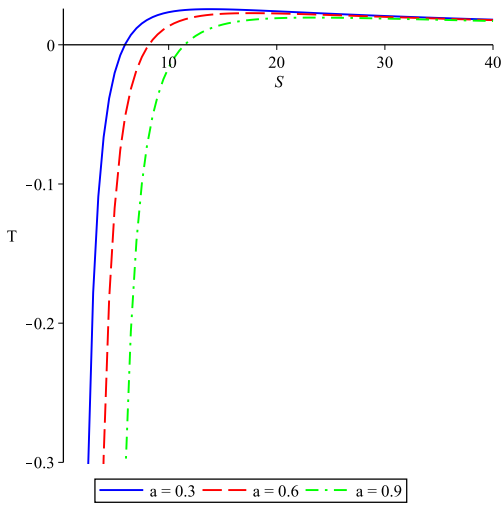}
	\caption{Variation of the black hole temperature $T$ in the presence of
		quintessence dark energy with characteristics $(Q,c,\epsilon)=(1,0.02, -\frac{2}{3})$. Here $a=0.3$ corresponds to blue continuous line, $a=0.6$ to red dash and $a=0.9$ to green dash-point.}
\end{figure}

To find out the impact of the absolute negative temperature, we have plot the variation of the mass $M$.
Ghaffarnejad et \textit{al.}\cite{ghaffarnejad2019reissner} putted out that absolute negative temperature could appear on Reissner-Nordström black hole, and could be caused by the decrease of entropy. However, the entropy of rotating and non-linear magnetic-charged black hole increases.

In FIG. 3, the mass of the black hole is plotted, and we can see that it decreases, but after reached at a minimum, starts increasing. Moreover, this minimum is shifted to higher radius values $r$, as we increase the magnetic charge $Q$. Afterwards, analyzing both FIG. 2 and FIG. 3,  we see that  absolute negative values of temperature could be caused by the decrease of black hole mass indeed.

However, FIG. 4 shows us that for high values of entropy, the black hole reaches a second decrease in its mass (FIG. 4.a), but this does not appear without quintessence parameter $c$(FIG. 4.b). This means that quintessence energy could affect the behavior of the black hole, in order to decrease its mass for higher values of horizon radius.
%


Afterwards, let us find the potential, through the Eq.(11).

We have
\begin{equation}
\Phi=\frac{3Q^2}{2}\left(\frac{1}{\frac{S}{\pi}-a^2}+\frac{a^2}{\left(\frac{S}{\pi}-a^2\right)^2}-\frac{c}{\left(\frac{S}{\pi}-a^2\right)^\frac{3\epsilon+2}{2}}\right).
\end{equation}

%
At the same time, Eq.(12) gives us
\begin{equation}
C=-\frac{Q^3+\left(\frac{S}{\pi}-a^2\right)^{\frac{3}{2}}}{2\left(\frac{S}{\pi}-a^2\right)^{\frac{3(\epsilon+1)}{2}}}.
\end{equation}

\begin{widetext}

	\begin{figure}[!h]
		\centering
		\begin{minipage}[t]{7cm}
			\centering
			\includegraphics[scale=0.35]{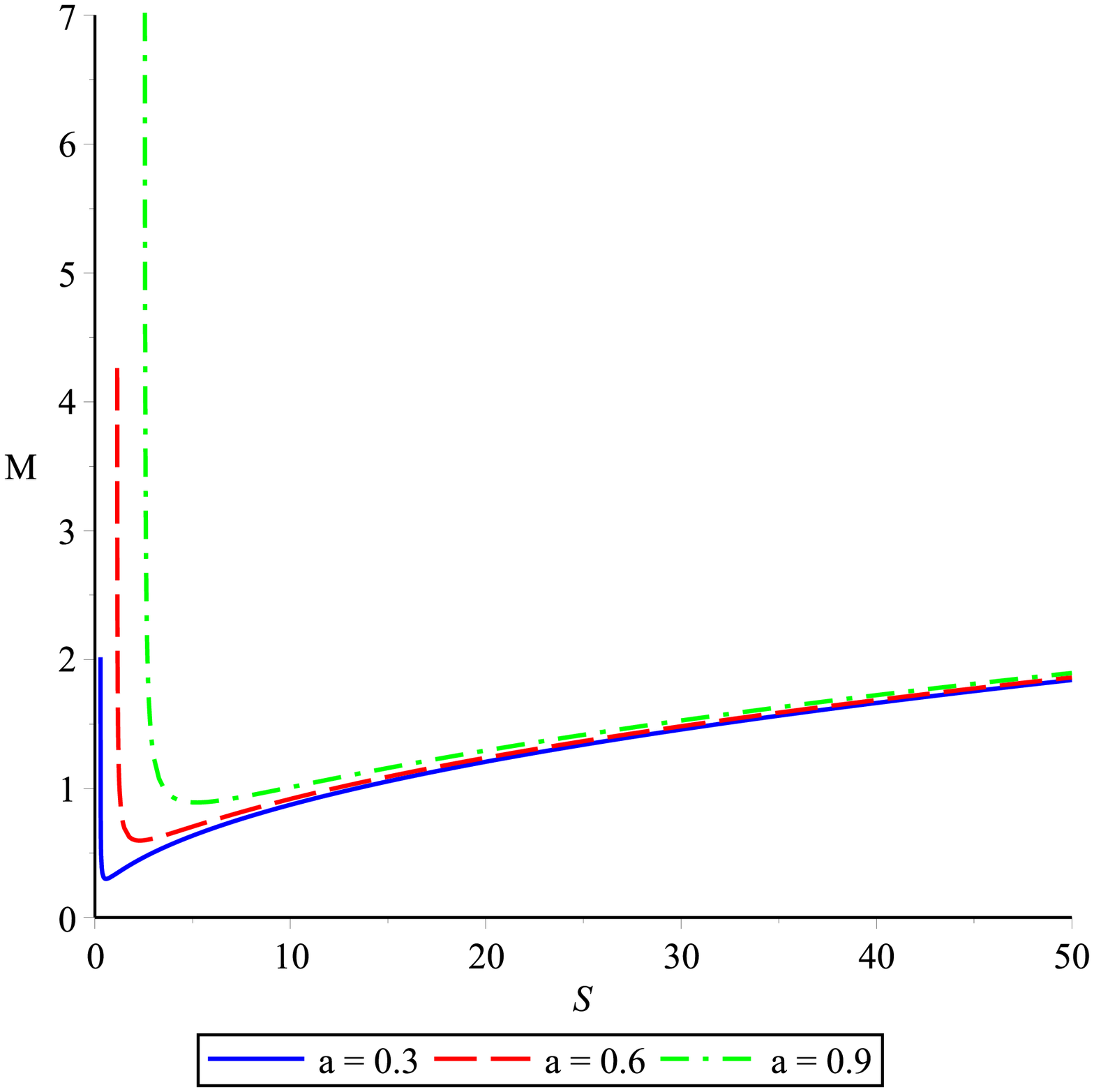}
			(a) for $Q=0$
		\end{minipage}
		\hspace{3cm}
		\begin{minipage}[t]{7cm}
			\centering
			\includegraphics[scale=0.35]{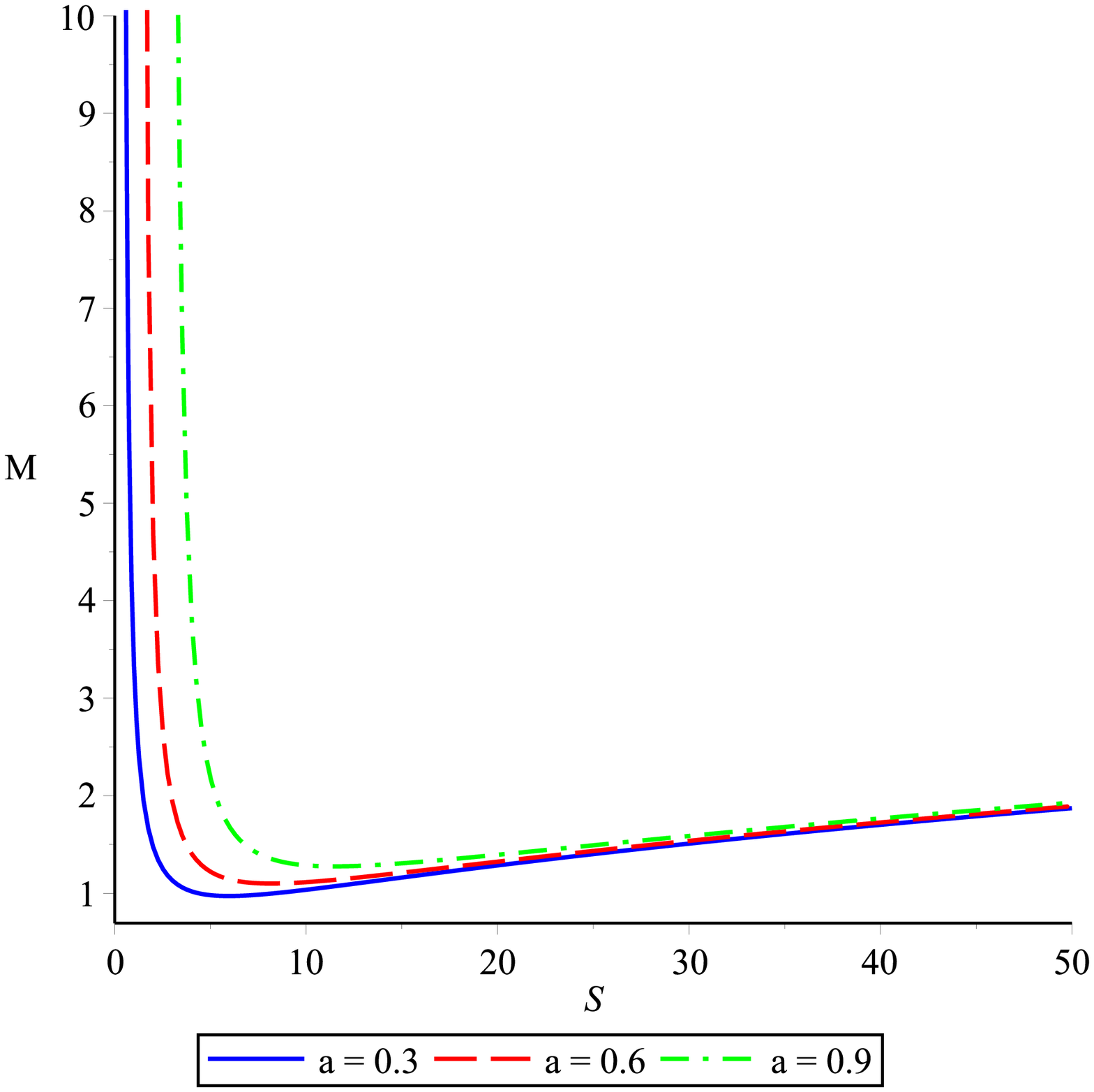}
			(b) for $Q=1$
		\end{minipage}
		\begin{minipage}[t]{7cm}
			\centering
			\includegraphics[scale=0.35]{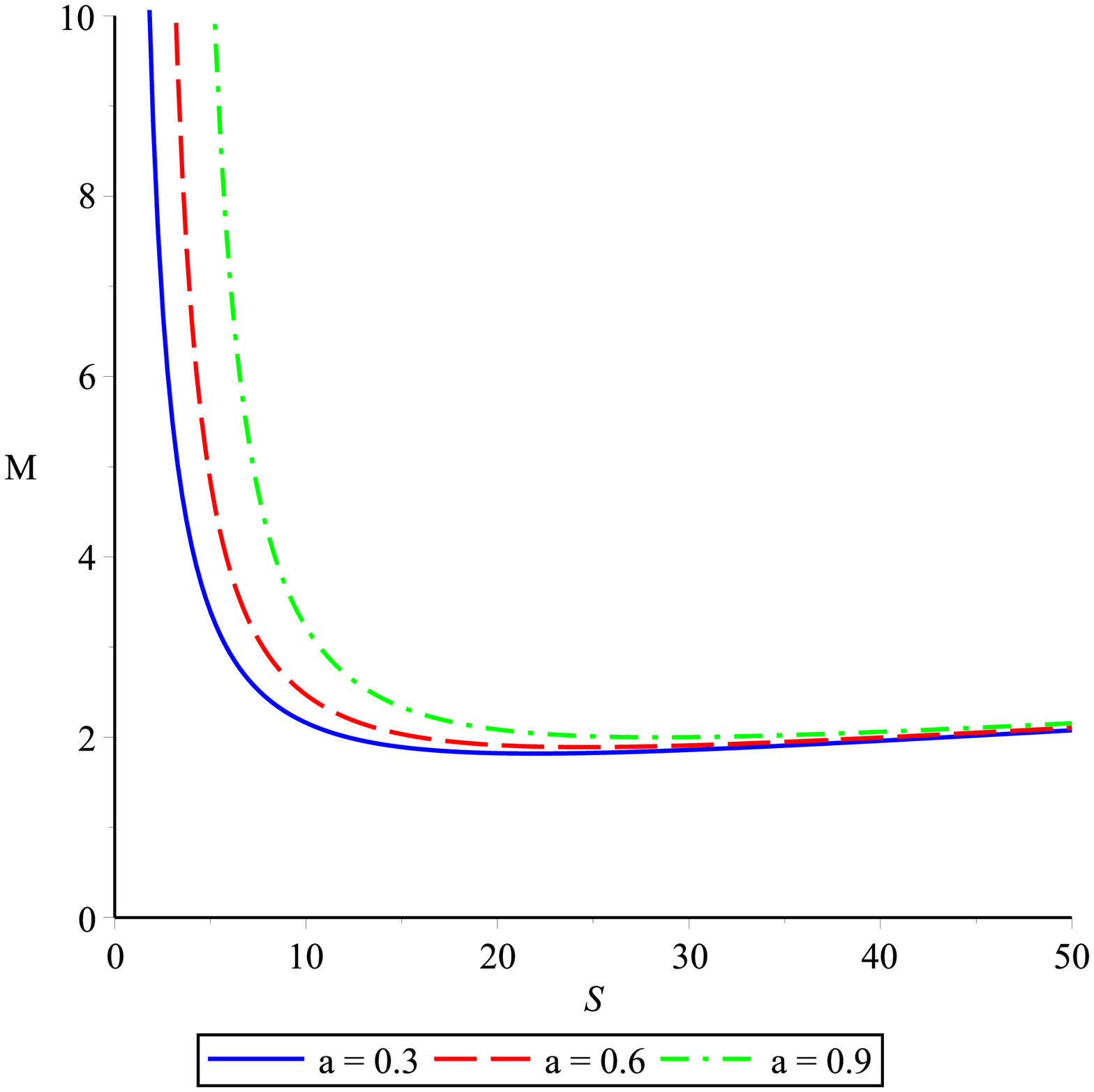}
			(c) for $Q=2$
		\end{minipage}
		\hspace{3cm}
		\begin{minipage}[t]{7cm}
			\centering
			\includegraphics[scale=0.35]{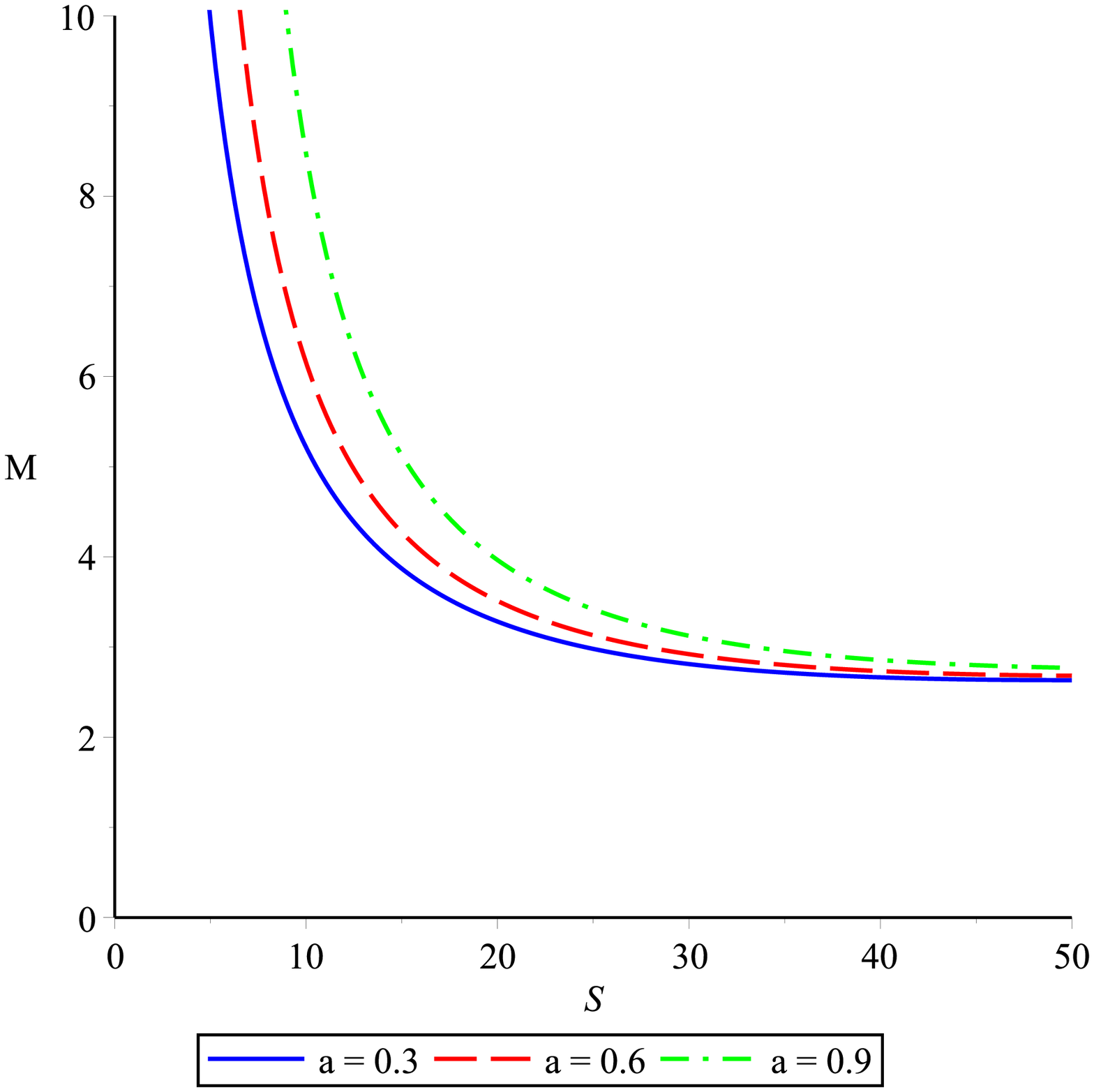}
			(d) for $Q=3$
		\end{minipage}
		\caption{Variation of the black hole mass $M$ in the presence of quintessence dark energy with characteristics $(c,\epsilon)=(0.02,-\frac{2}{3})$. Here, $a=0.3$ corresponds to blue continuous line, $a=0.6$ to red dash and $a=0.9$ to green dash-point.}
	\end{figure}
\end{widetext}

\begin{widetext}
	
	\begin{figure}[!h]
		\begin{minipage}[t]{7cm}
			\centering
			\includegraphics[scale=0.33]{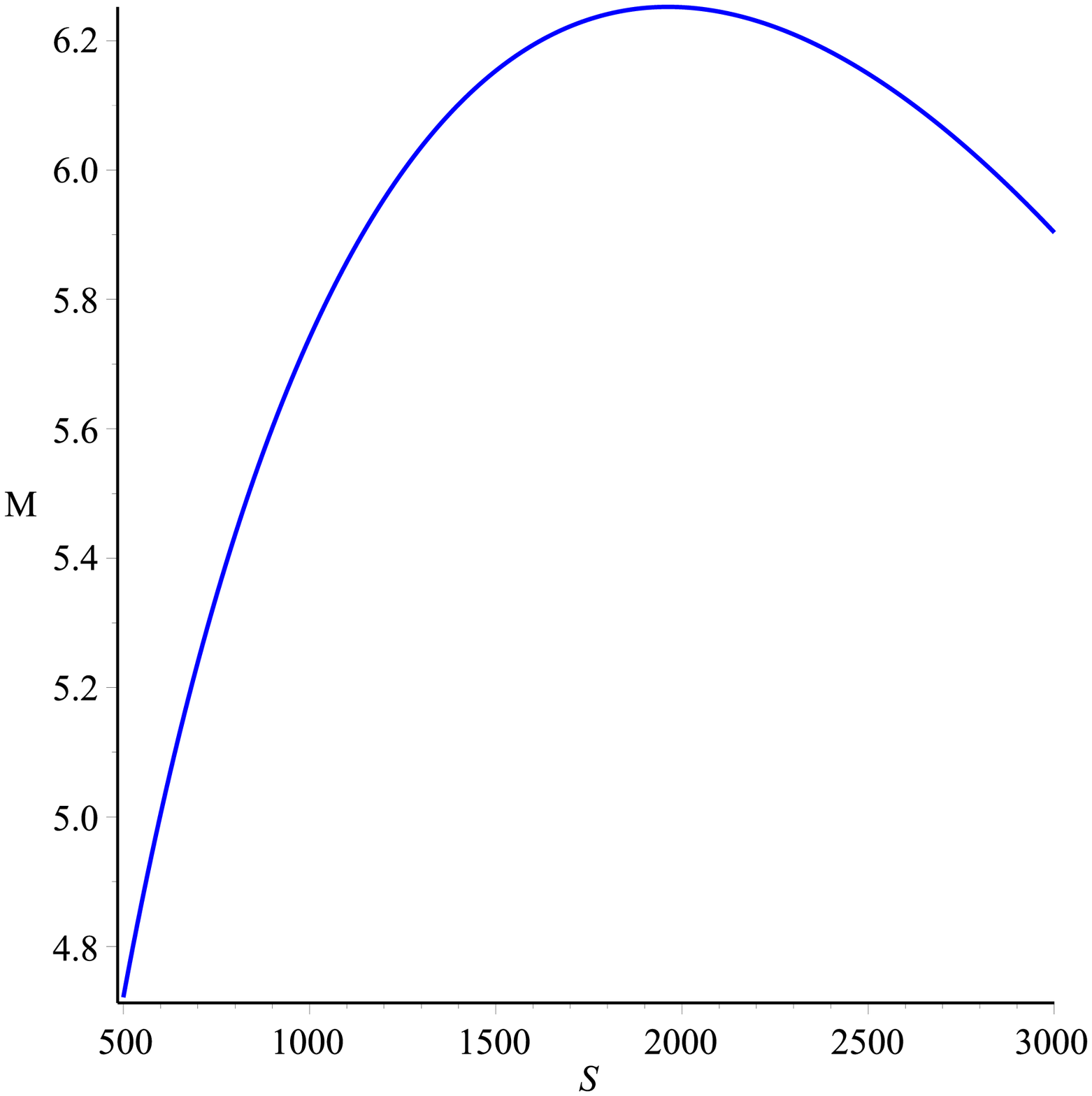}
			(a) for $c=0.02$
		\end{minipage}
		\hspace{3cm}
		\begin{minipage}[t]{7cm}
			\centering
			\includegraphics[scale=0.33]{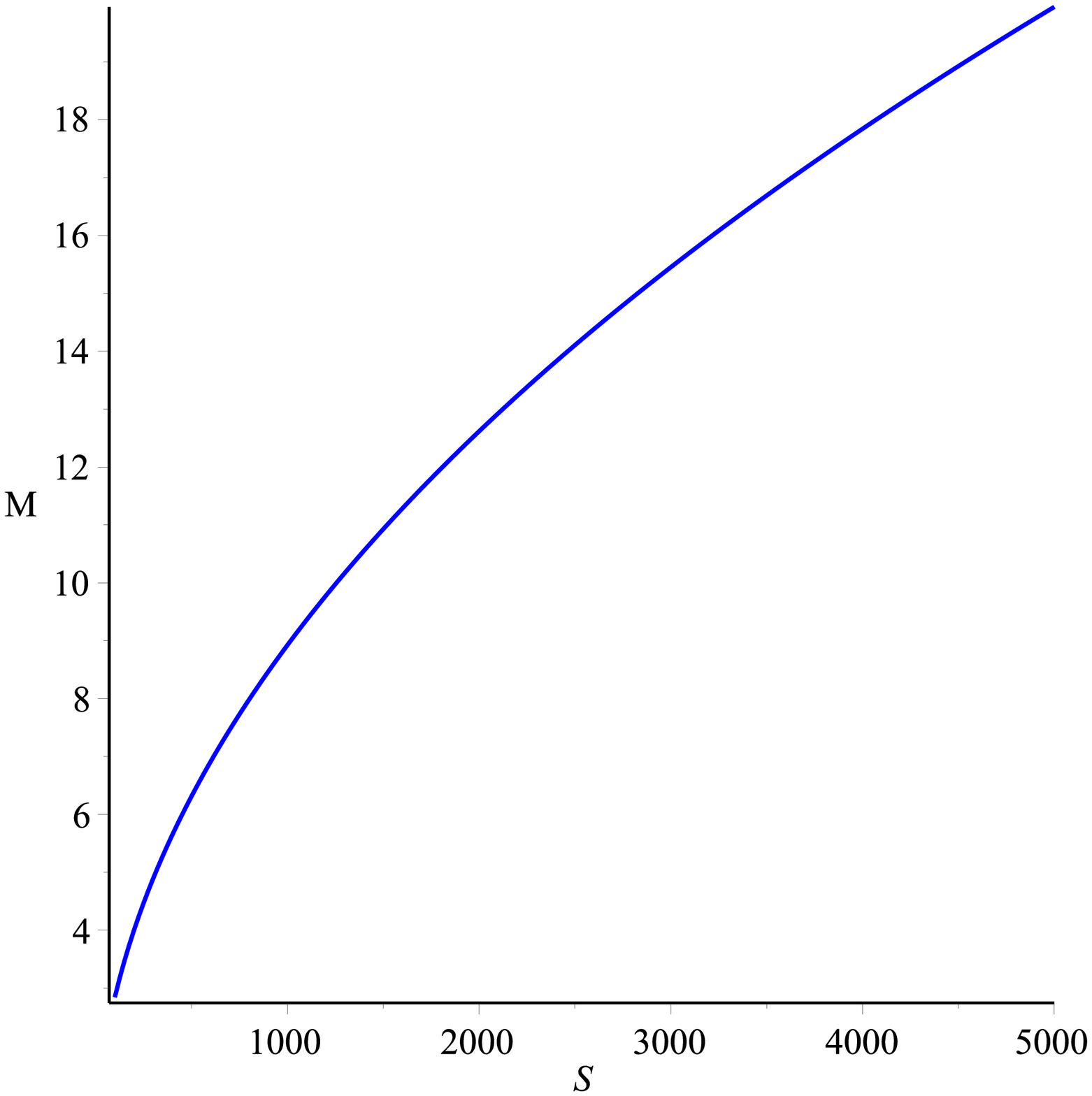}
			(b) for $c=0$
		\end{minipage}
		\caption{Variation of the black hole mass $M$ in the presence of
			quintessence dark energy with characteristics $(Q,a)=(1,0.3)$.}
	\end{figure}
\end{widetext}


%
\begin{center}
	\textbf{2. Phase transition}
\end{center}

Now, we are going to find the expression of heat capacity, which will permit us to determine at which conditions the black hole could be stable.


The formula necessary to compute the heat capacity is\cite{davies1977thermodynamic}
\begin{equation}
Ca_x=T\left(\frac{\partial S}{\partial T}\right)_x,
\end{equation}
with some set of parameters, denoted by $x$, held constant.
\newpage

To find out if a second-order phase transition occurs, we first have to remember the relation between the heat capacity and the free enthalpy, and then plot it. For that, the best thermodynamic function adapted to the study of these transformations is then the free enthalpy $G$ defined by
\begin{equation}
dG=-SdT+Vdp \ \ \ \  S=-\left.\frac{\partial G}{\partial T}\right|_V.
\end{equation}
Therefore, we have
\begin{equation}
C=-T\frac{\partial^2G}{\partial T^2}.
\end{equation}

Since $C$ is the second derivative of the free enthalpy, we have to study the curve of heat capacity and to detect if we have any presence of  first-order or second-order phase transition. These phenomena appear when heat capacity changes its sign without or with a discontinuity, respectively\cite{Mahamat2018}.
On the other hands, if we have a null constant magnetic charge $Q$, we get easily
\begin{equation}
Ca_{c,a}=2(S-\pi a^2)\frac{2c(S-\pi a^2)^{\frac{3}{2}}-\sqrt{\pi}(S-2\pi a^2)}{\sqrt{\pi}(S-4\pi a^2)}.
\end{equation}

This result has been also obtained by Toshmatov et \textit{al.}\cite{Toshmatov2017rotating}.

%
%

\begin{widetext}

	\begin{figure}[!h]
		\centering
		\begin{minipage}[t]{7cm}
			\centering
			\includegraphics[scale=0.34]{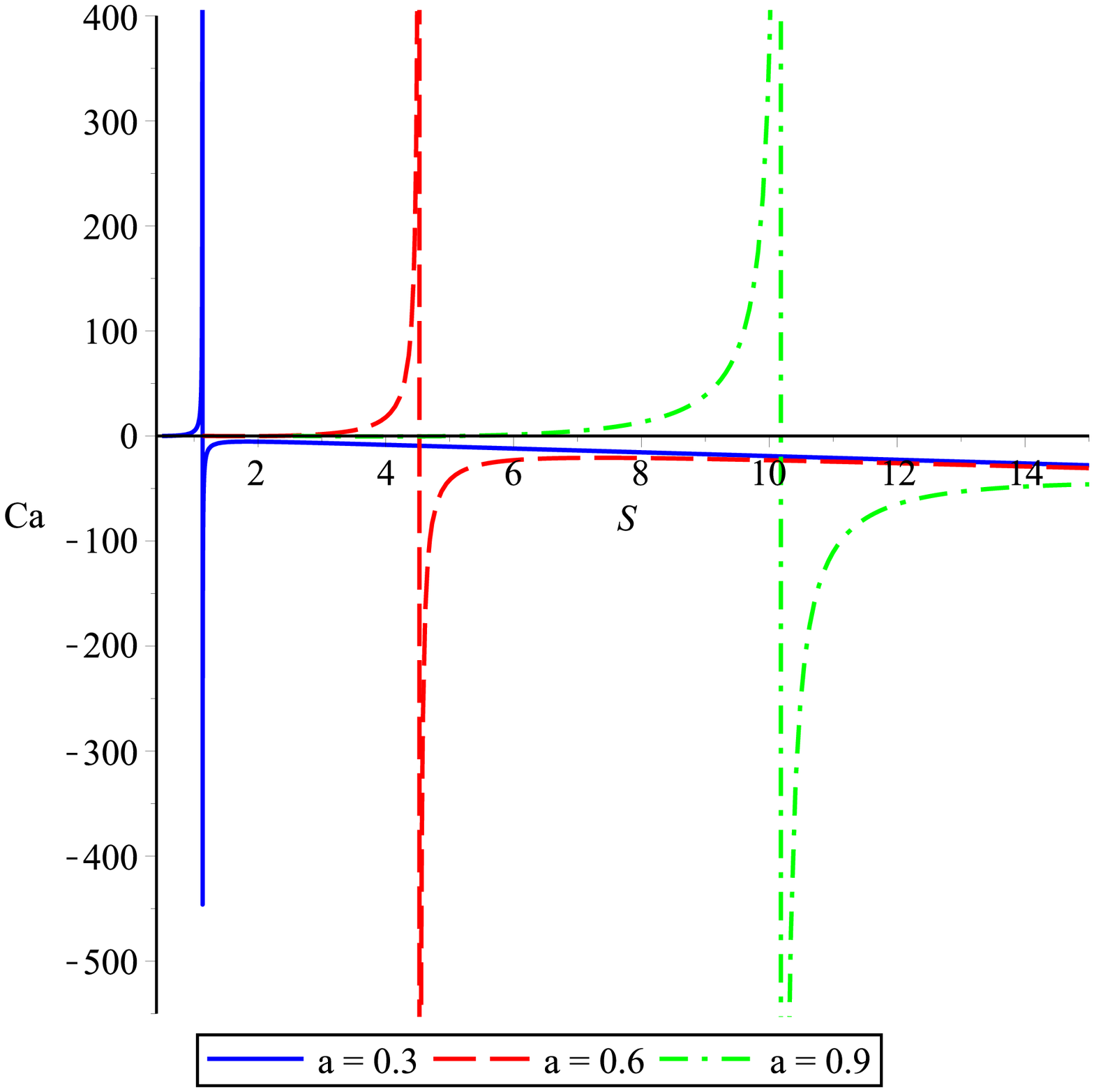}
			(a) for $Q=0$
		\end{minipage}
		\hspace{3cm}
		\begin{minipage}[t]{7cm}
			\centering
			\includegraphics[scale=0.34]{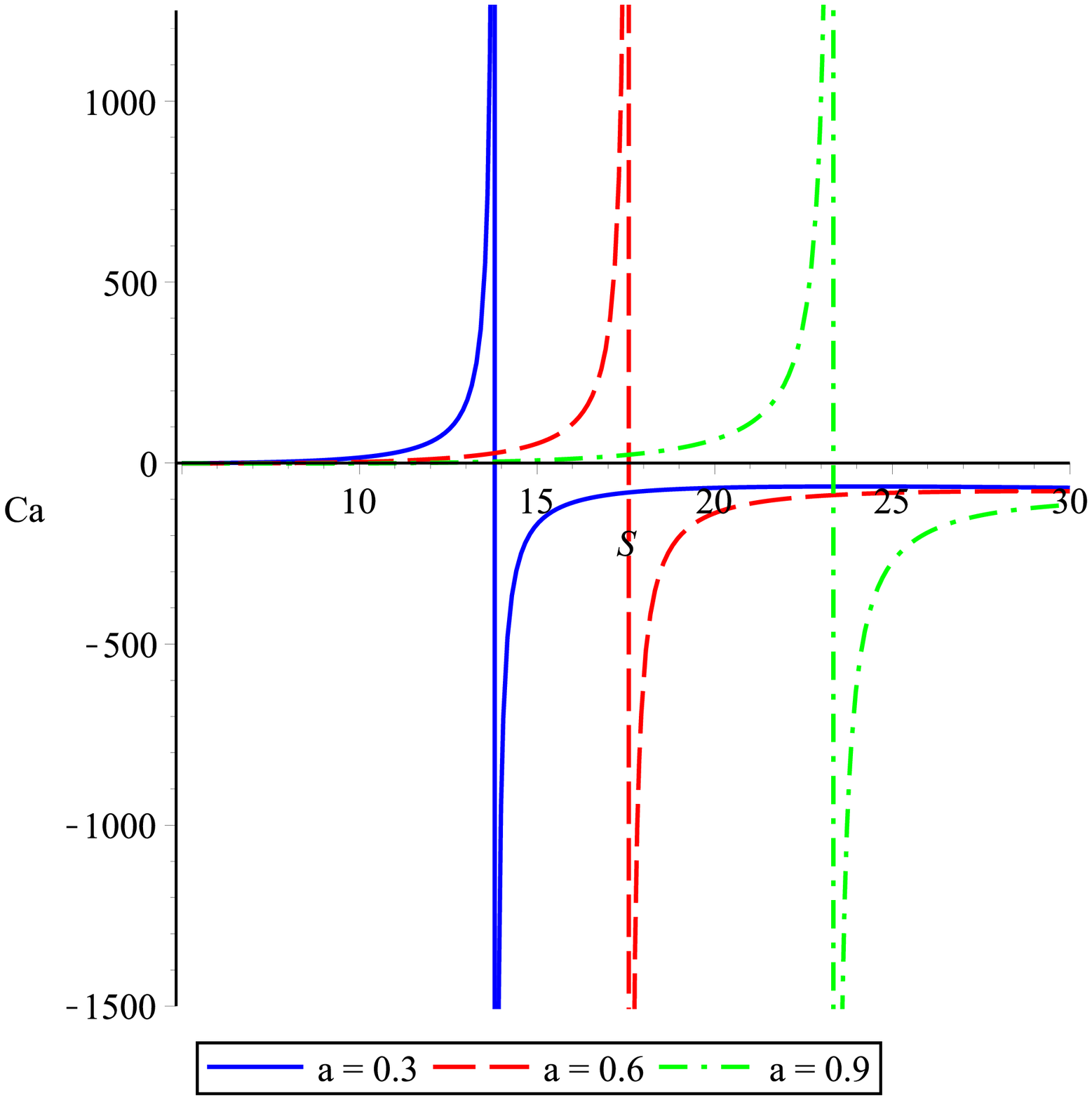}
			(b) for $Q=1$
		\end{minipage}
		\begin{minipage}[t]{7cm}
			\centering
			\includegraphics[scale=0.34]{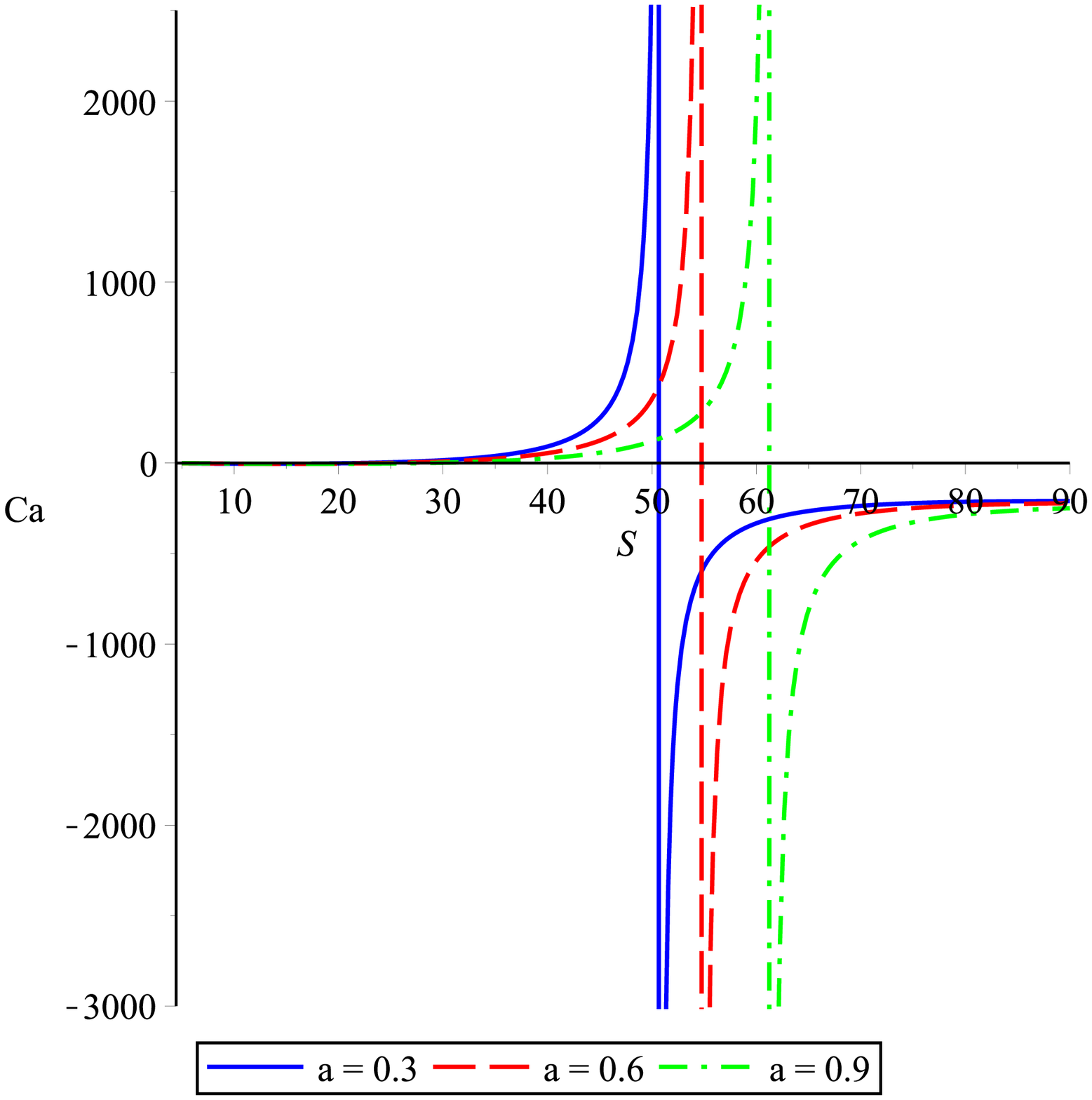}
			(c) for $Q=2$
		\end{minipage}
		\hspace{3cm}
		\begin{minipage}[t]{7cm}
			\centering
			\includegraphics[scale=0.34]{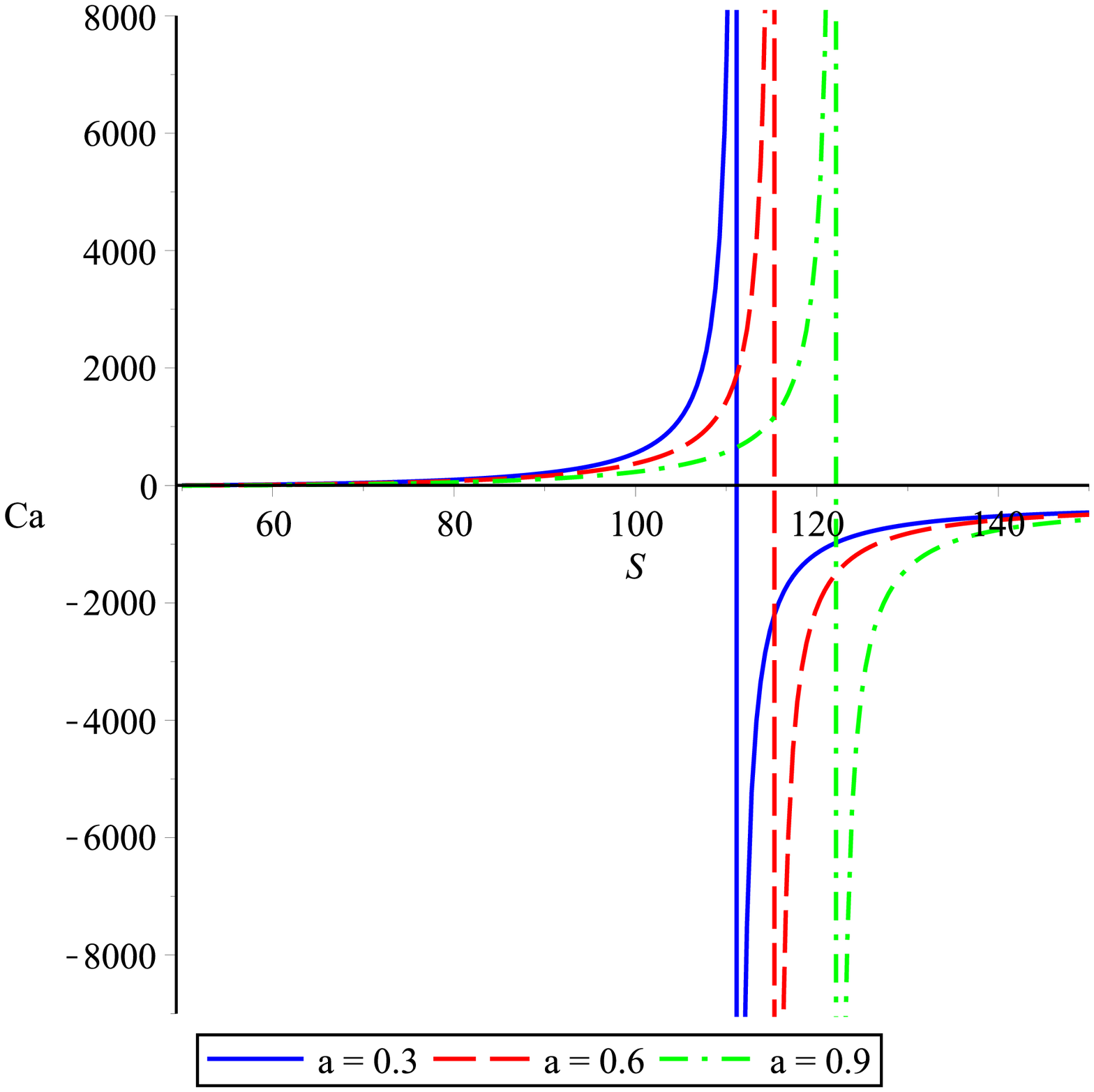}
			(d) for $Q=3$
		\end{minipage}
		
		\begin{minipage}[t]{7cm}
			\centering
			\includegraphics[scale=0.34]{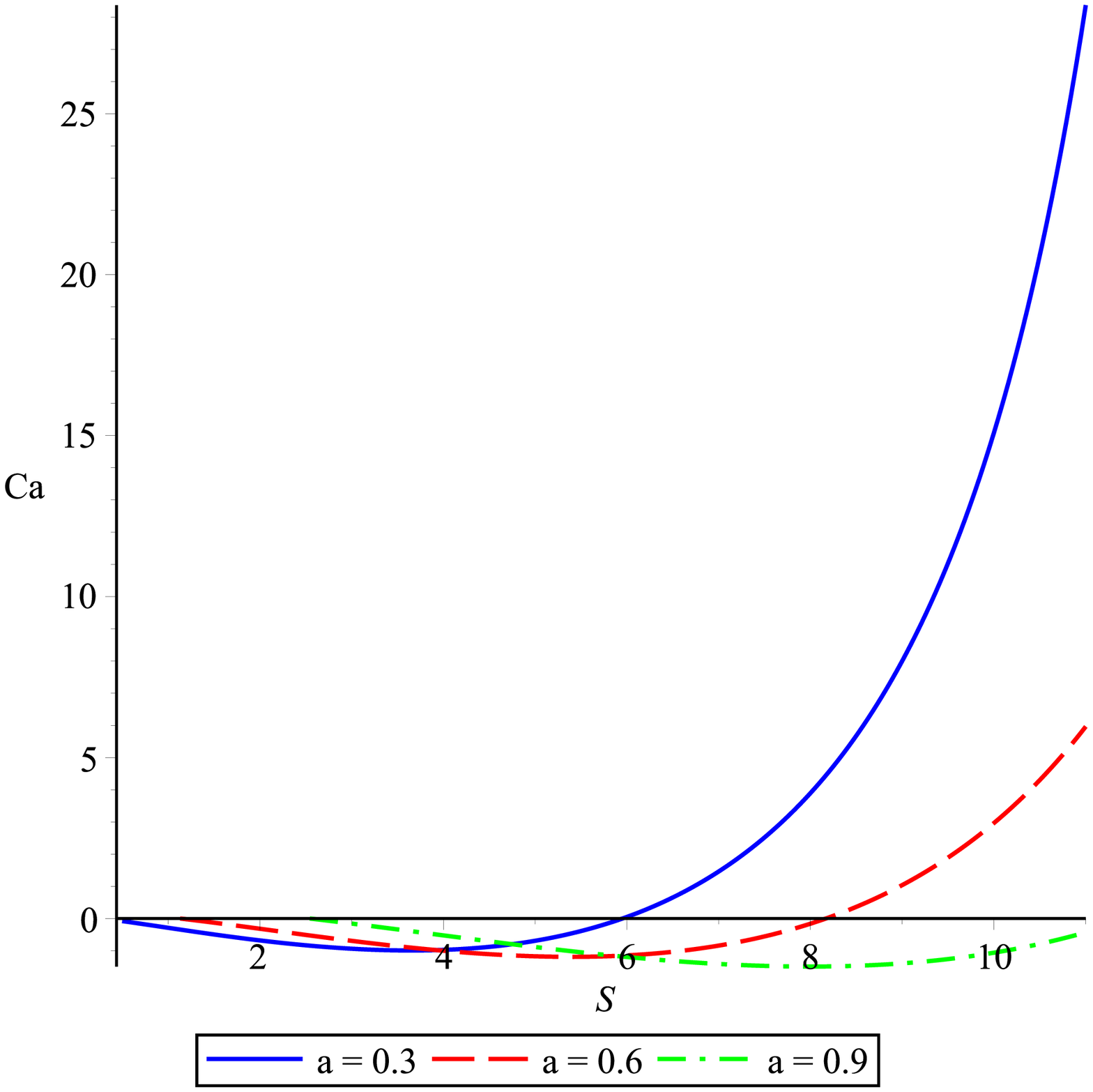}
			(e) for $Q=1$
		\end{minipage}
		\caption{Variation of the black hole heat capacity $Ca$ in the presence of quintessence dark energy with characteristic $(c,\epsilon)=(0.02,-\frac{2}{3})$. Here, $a=0.3$ corresponds to blue continuous line, $a=0.6$ to red dash and $a=0.9$ to green dash-point. (e) corresponds to lower values of entropy for $Q=1$.}
	\end{figure}
\end{widetext}
In FIG. 5, is plotted the heat capacity of rotating and non-linear magnetic-charged black hole with quintessence at constant magnetic charge $Q$, parameter $c$ and the rotating parameter $a$.

Analyzing these figures we can make a good appreciation of the black hole behavior. Precisely, we notice that a second-order phase transition occurs, localized by the presence of a discontinuity onto the  plot of the heat capacity. Precisely, analyzing FIG. 5(a), 5(b), 5(c) and 5(d), we notice that the black hole, being stable($Ca>0$), become unstable($Ca<0$) after this second-order phase transition. Moreover, we notice that the phase transition is shifted towards the higher values of entropy as we increase the rotating parameter $a$ or the magnetic parameter $Q$.

On the other hand, if we choose $Q=1$, and then make a comparison between FIG. 5(b) and 5(e) both corresponding to $Q=1$, we notice that for lower values of the entropy FIG. 5(e), we have a heat capacity which is negative, meaning that the black hole is unstable, but becomes positives, then being stable, through a first-order phase transition. But it becomes unstable again after the second-order phase transition for higher values of the entropy, as we can see in FIG. 5(e).

\newpage
\section{CONCLUSION}

In summary, we have studied the thermodynamic of rotating and non-linear magnetic-charged black hole in the quintessence field. First, we have found the entropy of such a black hole. Then, plotting the expression of this entropy, we have seen that this result could permit us to confirm the second law of black hole thermodynamic, which tels us that entropy $S$ of black hole is always increasing.

On the other hand, we have computed and plotted some thermodynamic quantities such as mass, temperature, and the potential provided from the magnetic charge, corresponding to this black hole. These results showed that, in term of entropy, the temperature increases from negatives values but reaches a maximum, before slowly decreasing with positive values.

Afterwards, we have found the expression of heat capacity of this kind of black hole. We plotted it and after analysis, it follows that the black hole undergoes to a first order phase transition for lower values of entropy;  afterwards, for higher values, we see a second-order phase transition. Moreover, this phase transition is shifted towards the higher values of entropy as we increase the rotating parameter $a$ or the magnetic parameter $Q$.


\end{document}